\documentclass[aps,showpacs,preprintnumbers,amssymb,superscriptaddress,twocolumn]{revtex4}

\usepackage{graphicx,color}
\usepackage{dcolumn}
\usepackage{bm}
\usepackage[active]{srcltx}

\begin{document}
\title{Fragile balance of the exchange interactions in Mn$_{1-x}$Co$_{x}$Ge compounds}

\author{E. Altynbaev}
\affiliation{Petersburg Nuclear Physics Institute, Gatchina, 188300 St-Petersburg, Russia}
\affiliation{Faculty of Physics, Saint-Petersburg State University, 198504 Saint Petersburg, Russia}
\author{S.-A. Siegfried}
\affiliation{German Engineering Materials Science Centre (GEMS) at Heinz Maier-Leibnitz Zentrum (MLZ), Helmholtz-Zentrum Geesthacht GmbH, Lichtenbergstr. 1, 85747 Garching bei M\"unchen, Germany}
\author{P. Strau\ss}
\affiliation{Institut f\"ur Physik der Kondensierten Materie, Technische Universit\"at Braunschweig, 38106 Braunschweig, Germany}
\author{D. Menzel}
\affiliation{Institut f\"ur Physik der Kondensierten Materie, Technische Universit\"at Braunschweig, 38106 Braunschweig, Germany}
\author{A. Heinemann}
\affiliation{German Engineering Materials Science Centre (GEMS) at Heinz Maier-Leibnitz Zentrum (MLZ), Helmholtz-Zentrum Geesthacht GmbH, Lichtenbergstr. 1, 85747 Garching bei M\"unchen, Germany}
\author{L. Fomicheva}
\affiliation{Institute for High Pressure Physics,  142190, Troitsk, Moscow Region, Russia}
\author{A. Tsvyashchenko}
\affiliation{Institute for High Pressure Physics,  142190, Troitsk, Moscow Region, Russia}
\author{S. Grigoriev}
\affiliation{Petersburg Nuclear Physics Institute, Gatchina, 188300 St-Petersburg, Russia}
\affiliation{Faculty of Physics, Saint-Petersburg State University, 198504 Saint Petersburg, Russia}

\begin{abstract}

The magnetic system of the pseudobinary compound Mn$_{1-x}$Co$_{x}$Ge has been studied using small-angle neutron scattering and SQUID-measurements. It is found that Mn$_{1-x}$Co$_{x}$Ge orders magnetically at low temperatures in the whole concentration range of $x \in [0 \div 0.9]$. Three different states of the magnetic structure have been found: a short-periodic helical state at $x \leq 0.45$, a long-periodic helical state at $0.45 < x \leq 0.8$, and a ferromagnetic state at $x \sim 0.9$. Taking into account that the relatively large helical wavevector $k \gg 1$ nm$^{-1}$ is characteristic for systems with mainly Ruderman-Kittel-Kasuya-Yoshida (RKKY) interaction, we suggest that the short-periodic helical structure at $x \leq 0.45$ is based on an effective RKKY interaction. Also the decay of $k$ with increasing $x$ is ascribed to a reduction of the interaction between second nearest neighbors and, therefore, to an increase of the influence of the Dzyaloshinskiy-Moriya interaction (DMI). As a result of the competition between these two interactions the quantum phase transition from a long-range ordered (LRO) to a short-range ordered (SRO) helical structure has been observed upon increase of the Co-concentration at $x_{c1} \sim 0.25$. Further increase of $x$ leads to the appearance of a double peak in the scattering profile at $0.45 < x < 0.7$. The transition from a helical structure to a ferromagnetic state found at $x = 0.9$ is caused by the weakening of DMI as compared to the cubic anisotropy. In summary, the evolution of the magnetic structure of Mn$_{1-x}$Co$_{x}$Ge with increasing $x$ is an example of a continuous transition from a helical structure based on the effective RKKY interaction to a ferromagnetic structure passing through a helical structure based on DMI.

\end{abstract}


\maketitle

\section{Introduction}

The cubic B20-type compounds (MnSi, FeGe, etc) are well known for their incommensurate magnetic structures with a very long period. The helical spin structure in these compounds is caused by the antisymmetric Dzyaloshinskii-Moriya interaction (DMI) due to a non-centrosymmetric arrangement of magnetic atoms \cite{Ishikawa, Lebech89, Kataoka, Bak}. The Ge-based B20 compounds (Mn$_{1-x}$Fe$_{x}$Ge, Mn$_{1-x}$Co$_{x}$Ge, Fe$_{1-x}$Co$_{x}$Ge, etc.) enhance the variety of magnetic properties compared to the silicides in terms of relatively high phase transition temperatures $T_C$ (up to 278 K for FeGe \cite{Tokura_nano_2,Grigoriev13PRL,Lebech89}) and a large value of the magnetic wavevector $k$ (up to 2.2 nm$^{-1}$ for MnGe \cite{Kanazawa11,Kanazawa12,Alt_14,Alt_16Pov,Tokura_nano_2,Grigoriev13PRL,Makarova12,Mirebeau}). The compound MnGe also exhibits a magnetic order-disorder phase transition which is smeared over a wide temperature range starting with the appearance of helical fluctuations below 100 K and ending with ferromagnetic nano-regions at temperatures above $T_{SR} \sim 175$ K \cite{Alt_14,Mirebeau16}. The temperature evolution of the magnetic structure of Mn$_{1-x}$Fe$_{x}$Ge with $x < 0.45$ is found to be similar to pure MnGe \cite{Alt_16Pov,Alt_16PRB}. A comprehensive small-angle neutron scattering study of Mn$_{1-x}$Fe$_{x}$Ge with $x < 0.45$ revealed that the intrinsic instability, which has been observed earlier for pure MnGe, is intensified by Fe-doping. The increase of the Fe content $x$ leads to a quantum phase transition from a LRO to a SRO helical structure at $x_{c1} \approx 0.35$ \cite{Alt_16PRB}.

It is already well known that the chiral magnetic structure of MnSi and FeGe compounds caused by the presence of DMI in the system \cite{Lebech89,Grigoriev05PRB,Grigoriev10PRB}. However, measurements of the Hall effect in Mn$_{1-x}$Fe$_{x}$Si revealed the presence of effective RKKY interaction in formation of the magnetic structure \cite{Demishev_PRL_2015,RK,K,Y}. The differences of the magnetic properties of the MnGe-based compounds from MnSi and FeGe together with the theoretical predictions of the value of DMI constant made in \cite{Koretsune_arxiv_2015, Gayles_arxiv_2015,Kikuchi_arxiv_2016} imply, that the main interaction generating the magnetic structure in MnGe is an effective RKKY exchange \cite{Dmitrienko}. Inversely to the scenario proposed for Mn$_{1-x}$Fe$_{x}$Si, the substitution of Mn by Fe atoms in MnGe leads to an increase of the DMI constant reducing the coupling between the second nearest neighbors \cite{Alt_16PRB,Dmitrienko}. This approach also explains the decrease of the wavevector $k$ upon Fe-substitution in Mn$_{1-x}$Fe$_{x}$Ge \cite{Grigoriev13PRL,Tokura_nano_2}. 

Another intriguing feature observed during the study of the mixed compounds Mn$_{1-x}$Fe$_{x}$Ge and Fe$_{1-x}$Co$_{x}$Ge is the change of magnetic chirality upon mixing the two different magnetic atoms (Mn and Fe, or Fe and Co) \cite{Tokura_nano_2,Grigoriev13PRL,Grigoriev14PRB}. Theoretical investigations suggest to ascribe the flip of the helix chirality as function of $x$ to a change of sign of the DMI constant \cite{Koretsune_arxiv_2015, Gayles_arxiv_2015,Kikuchi_arxiv_2016}.

 \begin{figure*}
 \includegraphics[width=0.95\textwidth]{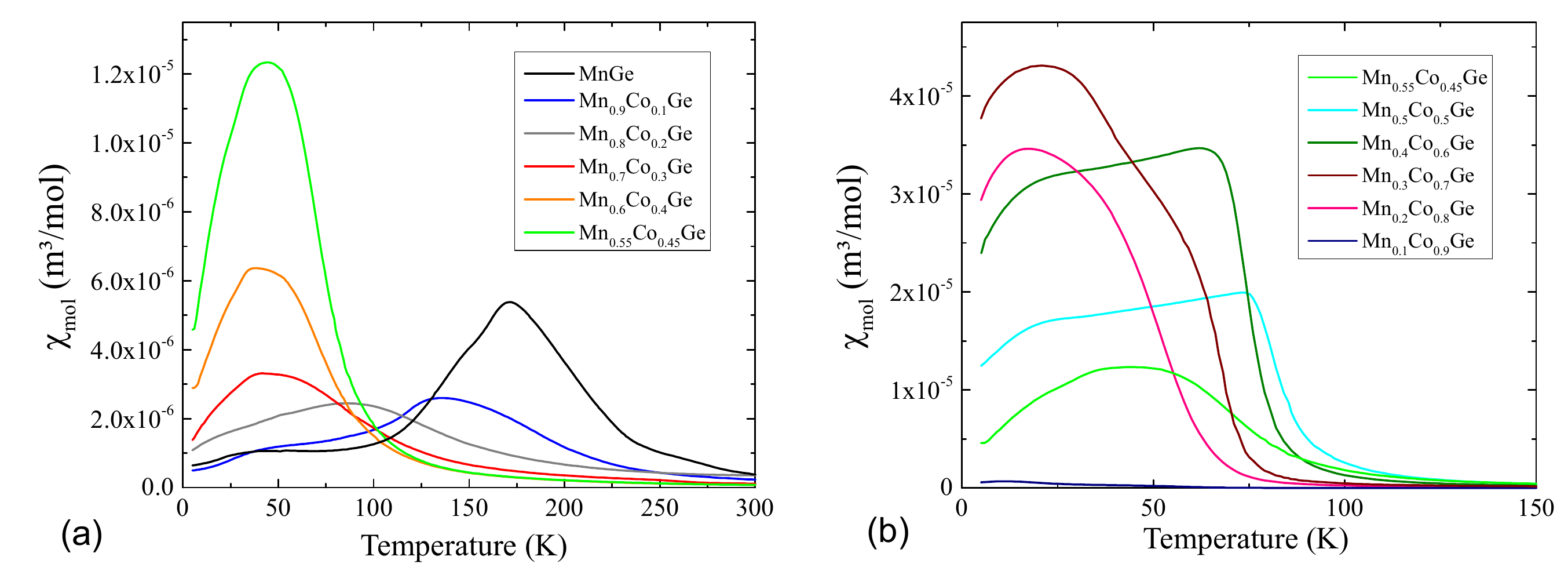}
 \caption{(color online) Temperature dependence of the magnetic susceptibility $\chi(T)$ of Mn$_{1-x}$Co$_{x}$Ge at $H = 10$ mT for (a) $x = 0.0$, 0.1, 0.2, 0.3, 0.4, 0.45 and (b) $x = 0.45$, 0.5, 0.6, 0.7, 0.8 and 0.9.}
 \label{fig:chi}
 \end{figure*}

Despite intensive research of the Ge-based B20 compounds, the Mn$_{1-x}$Co$_{x}$Ge solid solutions have not been studied yet. Similarly to the Mn$_{1-x}$Fe$_{x}$Ge compounds \cite{Alt_16PRB}, a quantum phase transition from the LRO to the SRO of the helical structure upon increase of Co concentration is expected. In accordance with the results of X-ray diffraction in combination with neutron diffraction \cite{Grigoriev13PRL,Grigoriev14PRB} as well as Lorentz transmission electron microscopy and electron diffraction \cite{Tokura_nano_2}, the Mn-based and Co-based transition metal monogermanides exhibit the same chiral connection between the crystalline handedness and the chirality of magnetic helix. Therefore, one possibility of the evolution of magnetic structure of Mn$_{1-x}$Co$_{x}$Ge compounds is the double flip of the chirality upon increase of $x$ in Mn$_{1-x}$Co$_{x}$Ge as long as such phenomena was observed in both series of the compounds: Mn$_{1-x}$Fe$_{x}$Ge and Fe$_{1-x}$Co$_{x}$Ge \cite{Grigoriev13PRL,Tokura_nano_2,Grigoriev14PRB}. Another option is the absence of the flip of the chiral link between structure and magnetism if the amount of 3d-electrons is not the only driving force for this effect. Nevertheless, a transition from a helimagnetic DMI-based structure to a ferromagnetic state due to the domination of the cubic anisotropy over the DMI in the compound \cite{Grigoriev15PRB} cannot be excluded.

In this work we present a comprehensive study of the changes of the magnetic properties of Mn$_{1-x}$Co$_{x}$Ge as function of the Co-substitution $x$ within the range $0.0 \leq x \leq 0.9$ using small-angle neutron scattering (SANS) together with SQUID measurements. We show that the quantum phase transition from the helical LRO to the SRO indeed takes place in Mn$_{1-x}$Co$_{x}$Ge at $x_{c1} \approx 0.25$. The critical Co content at which the DMI as main interaction dominates completely the effective RKKY interaction and causes the magnetic structure has been found as $x_{c2} \approx 0.45$. With further increase of $x$, a ferromagnetic state has been found for $x \geq 0.9$, which is ascribed to the weakening of DMI as compared to the cubic anisotropy. In summary, the evolution of the magnetic structure of the Mn$_{1-x}$Co$_{x}$Ge with increase of the Co content has been investigated and a continuous transition from a helical structure based on the effective RKKY interaction changing to a DMI-dominated helix which finally ends up in a ferromagnetic structure has been revealed.

\section{Sample preparation and magnetization}

Polycrystalline samples of Mn$_{1-x}$Co$_{x}$Ge have been synthesized by the high pressure method at the Institute for High Pressure Physics, Troitsk, Moscow, Russia. As they can be only synthesized under high pressure, the samples have a polycrystalline form with a crystallite size not less than $10-100$ microns (see \cite{Tsvyashchenko84} for details). X-ray powder diffraction confirmed the B20 structure of the samples used in experiments with an amount of impurities less than 2\% of volume fraction \cite{Chernyshov, Valkovskiy}.


Magnetic susceptibility measurements have been performed in order to obtain the magnetic order temperatures for all the compounds using a Quantum Design MPMS-5S SQUID magnetometer. The temperature dependent susceptibility has been measured upon heating in a magnetic field of $10$ mT after zero-field-cooling down to $T=5$ K. The susceptibility data obtained are presented in Fig.\ref{fig:chi} for Mn$_{1-x}$Co$_{x}$Ge compounds with $x = 0.0$, 0.1, 0.2, 0.3, 0.4, 0.45 (a) and $x = 0.45$, 0.5, 0.6, 0.7, 0.8, 0.9 (b). 

 \begin{figure*}
 \includegraphics[width=0.95\textwidth]{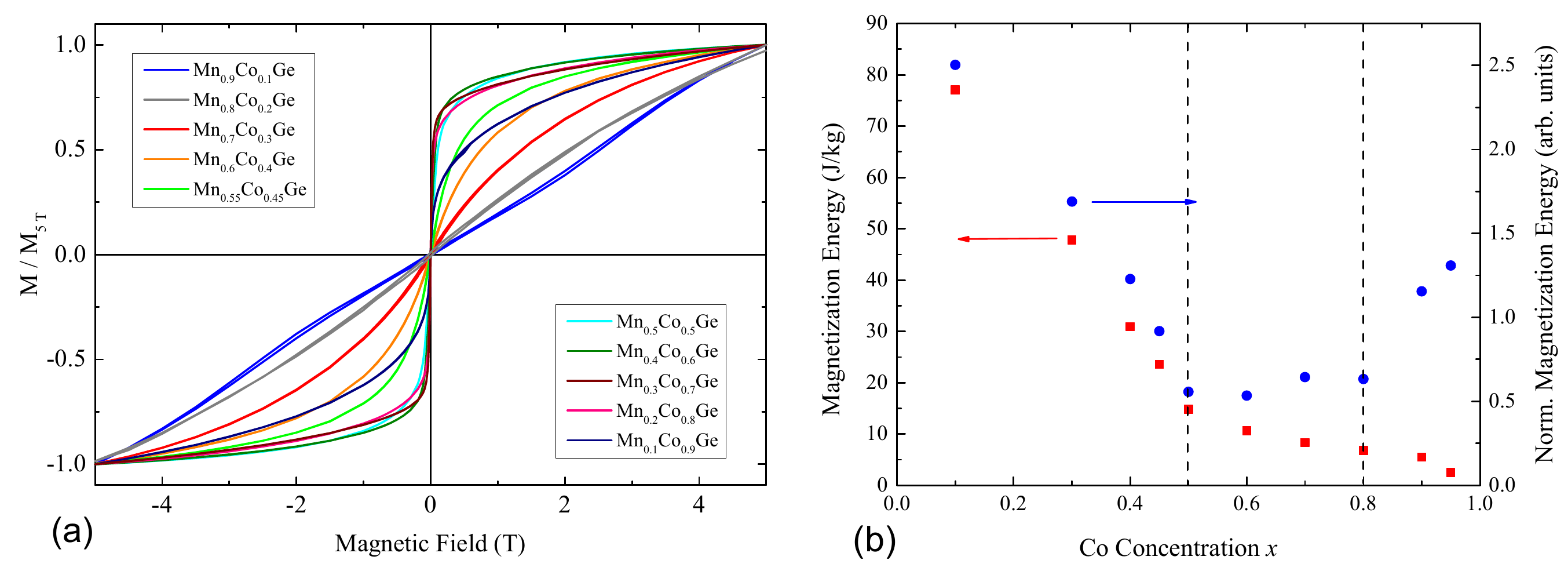}
 \caption{(color online) (a) Magnetization curves of Mn$_{1-x}$Co$_{x}$Ge at $T = 0.2 \cdot T_C$ normalized to the magnetization at $B = 5$~T. (b) Magnetization energy of Mn$_{1-x}$Co$_{x}$Ge derived from the data in (a) in absolute units (red squares) and taken from the normalized magnetization (blue circles).}
 \label{fig:hyst}
 \end{figure*}

The magnetization as function of the applied field has been measured at low temperature $T = 0.2 \cdot T_C$ far away from the magnetic order transition (Fig. \ref{fig:hyst}a). In the low range of Co concentration $x < 0.5$ the curves show a hard magnetic behavior which changes to a more magnetically anisotropic regime for $0.5 \leq x \leq 0.8$ where the compounds start to saturate at the highest available field ($B_{max} = 5$~T). For $x > 0.8$ the slope of the magnetization curves restarts to decrease upon Co substitution. For $x \leq 0.5$ and $x \geq 0.8$ the magnetization is still not saturated at 5~T which is for the low Co content in accordance to the behavior of pure MnGe \cite{Kanazawa11}. The slope of the magnetization curves reflects the magnetic anisotropy energy which is connected to the magnetization energy $W = \int{H \cdot dM}$. In Fig. \ref{fig:hyst}b the magnetization energy derived from the magnetizaton curves is plotted. The absolute energy values are directly calculated from the individual magnetization curve of each mixed compound. The normalized magnetization energy denotes the energy from the normalized magnetization $M/M_{5T}$ as shown in Fig. \ref{fig:hyst}a. From this curve three different regimes can be distinguished: (i) For $x \leq 0.5$ the magnetization energy is relatively high and decreases with increasing Co concentration. (ii) For $0.5 \leq x \leq 0.8$ the magnetization energy is low, which is correlated with a high magnetic anisotropy, and is nearly independent of the Co concentration. (iii) For $x > 0.8$ the normalized magnetization energy starts to increase again, whereas the absolute values become smaller due to the reduction of the magnetic moment at high Co concentration. This result already suggests that the exchange mechanism in the Mn$_{1-x}$Co$_{x}$Ge system changes upon Co substitution at $x \sim 0.5$ and $x \sim 0.8$ as indicated by the two dashed lines in Fig. \ref{fig:hyst}b.

\begin{figure}
 \includegraphics[width=0.45\textwidth]{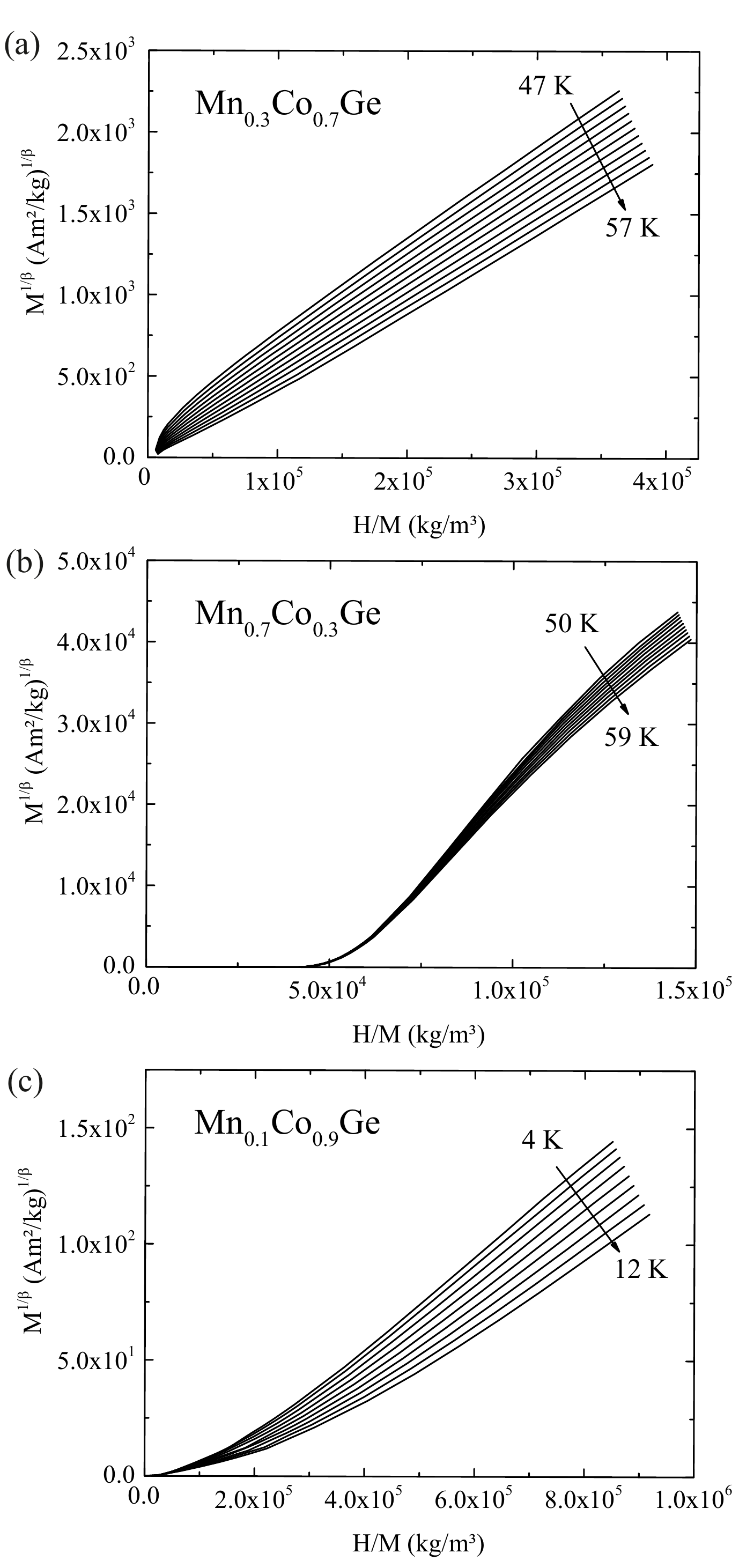}
 \caption{Arrott plots with $\beta = 0.31$ for (a) Mn$_{0.3}$Co$_{0.7}$Ge, (b) Mn$_{0.7}$Co$_{0.3}$Ge, and (c) Mn$_{0.1}$Co$_{0.9}$Ge.}
 \label{fig:arrott}
 \end{figure}

From isothermal magnetization curves Arrott plots have been obtained in order to determine the ordering temperature $T_C$. Only in the regime between $x = 0.5$ and $x = 0.7$ a linear behavior of $M^{1/\beta}$ vs. $(H/M)$ with $\beta = 0.31$ can be observed which allows for the identification of $T_C$ [Fig. \ref{fig:arrott}a for Mn$_{0.3}$Co$_{0.7}$Ge]. In the range $x < 0.5$ the field dependence of the magnetization changes in a way that the Arrott plots cannot be extrapolated to the origin of the diagram [Fig. \ref{fig:arrott}b for Mn$_{0.7}$Co$_{0.3}$Ge]. The reason for the deviation from linearity of the Arrott plots is most likely related to fluctuations close to the magnetic order transition which might also be of first order as known from MnSi \cite{Jaschonek}. Also for $x \geq 0.8$ the Arrott plots do not yield the ordering temperature [Fig. \ref{fig:arrott}c for Mn$_{0.1}$Co$_{0.9}$Ge]. That brings to the conclusion that the small-angle neutron scattering experiments are needed to investigate completely the microscopic nature of the magnetic states as function of the temperature for the different mixed compounds.

\section{SANS experiment}

 \begin{figure*}
 \includegraphics[width=0.95\textwidth]{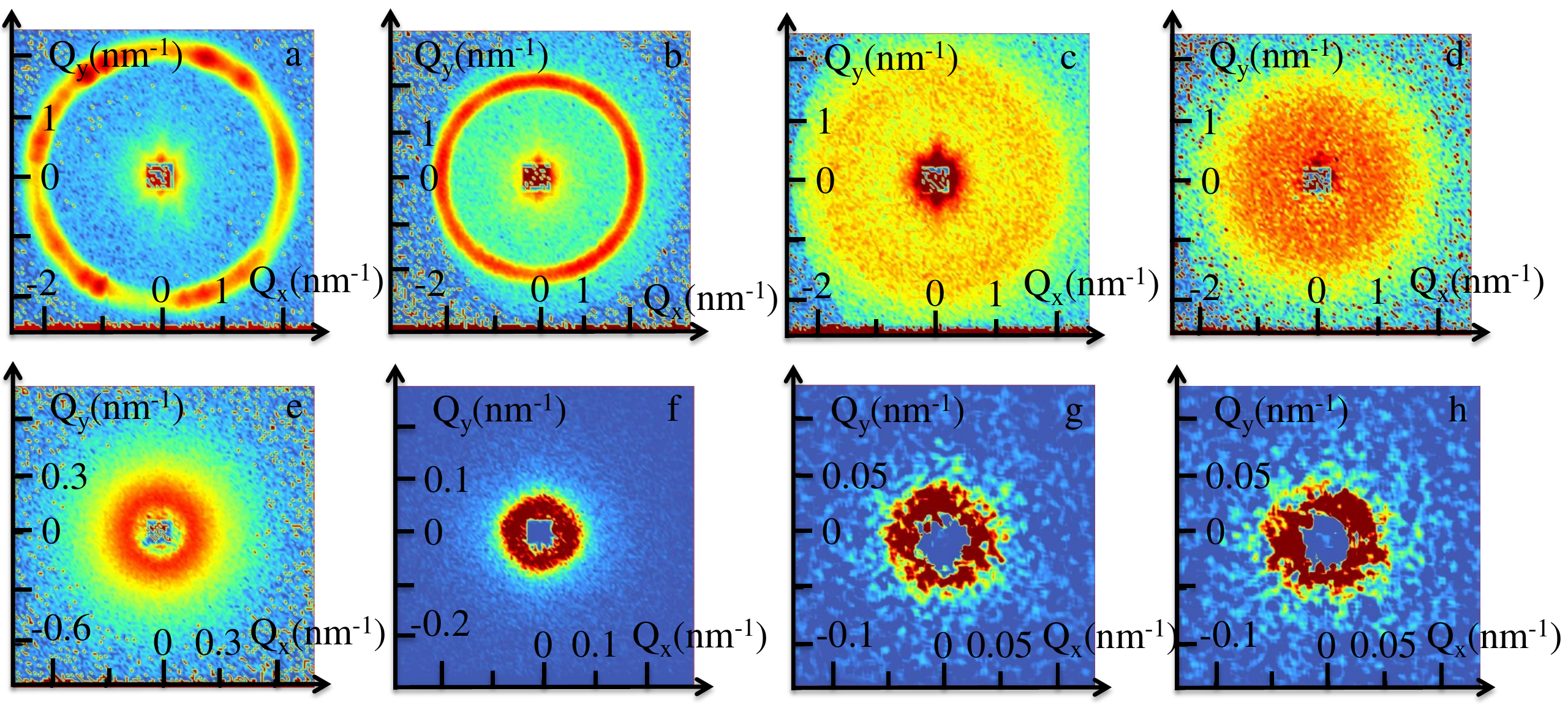}
 \caption{(color online) Examples of the neutron scattering maps for Mn$_{1-x}$Co$_x$Ge compounds with $x = 0.0$ (a), 0.1 (b), 0.3 (c), 0.4 (d), 0.5 (e), 0.6 (f), 0.7 (g) and 0.8 (h) at $T = 5$ K taken at zero field.}
 \label{fig:mapsx}
 \end{figure*}

 \begin{figure}
 \includegraphics[width=0.5\textwidth]{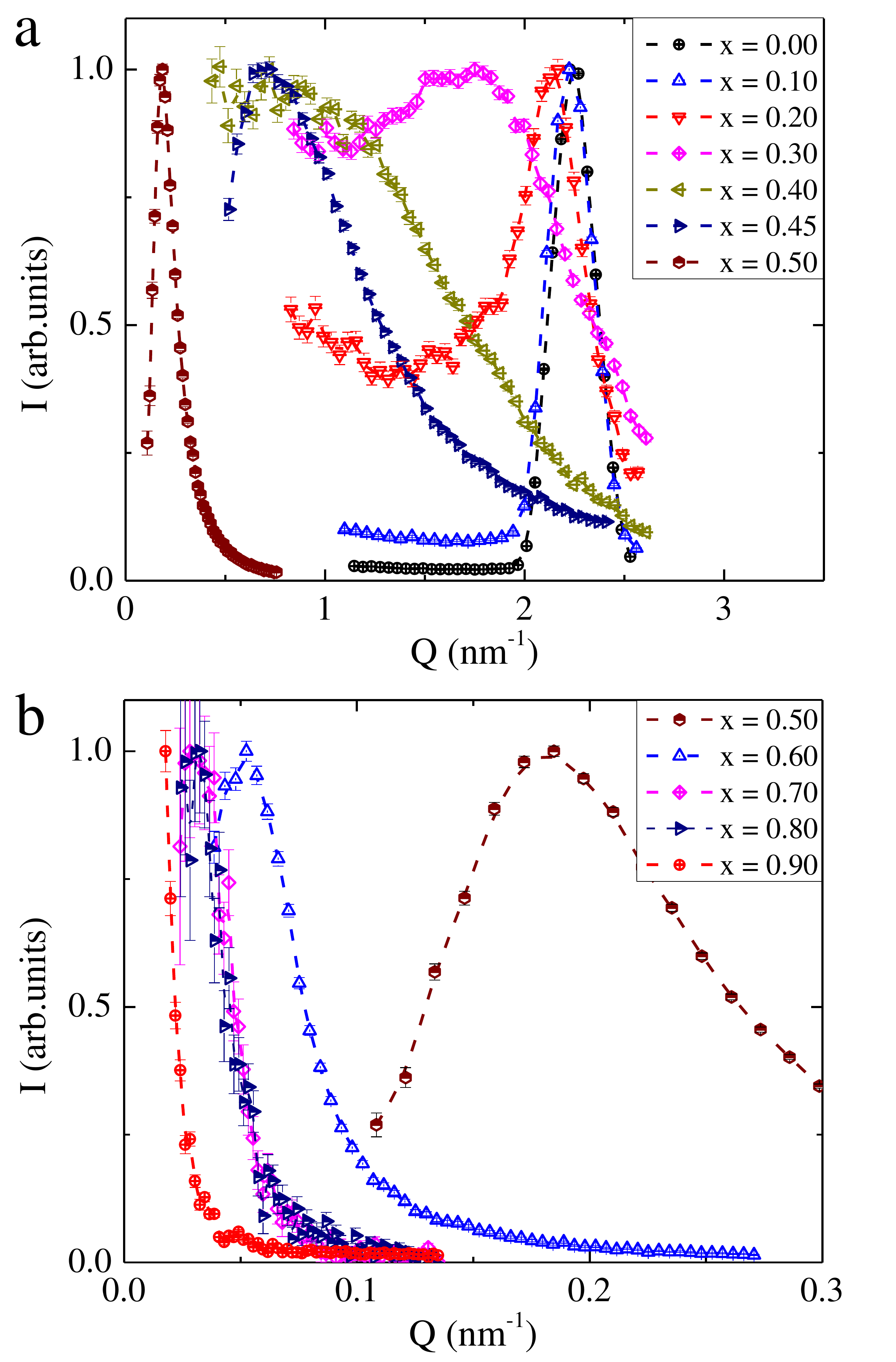}
 \caption{(color online) Momentum transfer dependence of the scattering intensity at $T = 10$ K for Mn$_{1-x}$Co$_x$Ge compounds with (a) $0.0 < x < 0.5$ and (b) $0.5 < x < 0.9$.}
 \label{fig:ivq}
 \end{figure}

In order to identify the magnetic structure of the compounds, SANS measurements were carried out using the SANS-1 instrument \cite{SANS} at the FRM-II reactor in Garching, Germany). Neutrons with a mean wavelength of $\lambda = 0.6$ nm and 1.2 nm were used. The sample--detector distance of 2, 7, and 20 meters was set to cover the range of momentum transfer $Q$ from $0.02$ nm$^{-1}$ to $2.7$ nm$^{-1}$ with the resolution equal to $20$ \%. The scattering intensity was measured upon zero-field cooling from the paramagnetic phase at $T = 300$ K to the ordered phase at $T = 5$ K.

Figure \ref{fig:mapsx}(a-h) shows examples of the small-angle neutron scattering maps for Mn$_{1-x}$Co$_x$Ge with $x$ from 0.0 to 0.8 at $T = 5$ K. The typical powder-like images were detected for the samples with $x = 0.0$ and 0.1 (Fig. \ref{fig:mapsx}a, b). From high resolution X-ray data we conclude that all samples have the same size of powder grains in the order of 10 microns \cite{Chernyshov,Valkovskiy}. With further increase of $x$ up to 0.4, the intensity of the ring smears over the large $Q$-area (Fig. \ref{fig:mapsx}c, d). This implies that the correlation length of the structure decreases dramatically. One should note that the evolution of the magnetic structure of Mn$_{1-x}$Co$_x$Ge with $x$ shows a surprising similarity to the evolution of the magnetic structure of pure MnGe with the temperature \cite{Alt_14}. For Co concentrations $x > 0.4$ the $k$-value decreases drastically (Fig. \ref{fig:mapsx}e-g). For the compound with $x = 0.9$ the value of $k$ is not detected within the resolution of the SANS instrument (Fig. \ref{fig:mapsx}g). Therefore this compound exhibits ferromagnetic order rather than a helical spin structure.

The scattering intensity $I({\bf Q})$ was azimuthally averaged and plotted for compounds with $0.0 < x < 0.5$ in Fig. \ref{fig:ivq}a and for compounds with $0.5 < x < 0.9$ in Fig. \ref{fig:ivq}b. The position of the Bragg reflection decreases to one-tenth of ${\bf Q}$ from $x = 0.0$ towards $x = 0.5$. In addition, the shape of the scattering profile changes dramatically in the same concentration range with increasing $x$ (Fig. \ref{fig:ivq}a). With further increase of Co concentration the value of the wave-vector $k$ tends to zero at $x = 0.9$. We present the detailed analysis of the SANS data separately for $0.0 < x < 0.45$ and $0.5 < x < 0.9$ because these two concentration ranges have to be discussed within two different approaches.

\section{RKKY-based helical structure}

Relatively large value of the helical wavevector, $ka \sim 1$, where $a$ is the lattice constant of the compound, \cite{Kanazawa11,Makarova12,Alt_14} together with the small predicted value of DMI constant \cite{Koretsune_arxiv_2015,Gayles_arxiv_2015,Kikuchi_arxiv_2016} allow us to consider the effective RKKY as the main driving force that built the magnetic structure of MnGe \cite{Dmitrienko}. The scattering function consists of two different parts: the inelastic contribution to the scattering function and the scattering from the helical magnetic order. The abnormal scattering at $Q < k$, well discussed in \cite{Alt_14} and considered as inelastic part of the scattering intensity can only be described by the step function convoluted with Lorentzian function. The intensity of this step-like scattering rises with $x$ at very low temperatures to its maximal value for $x = 0.4$ and then starts to decrease (Fig. \ref{fig:ivq}a). The Bragg reflection coming from the magnetic structure of Mn$_{1-x}$Co$_x$Ge compounds with $x < 0.45$ and can purely be described by a pseudo-Voigt function with four different parameters: the scaling factor $I_{Max}$, the Lorentz fraction $\alpha$, the peak position $k$ and the width of both, Gaussian and Lorentzian functions $\kappa$ (Fig. \ref{fig:ivq}a). The Lorentzian and Gaussian contribution to the Bragg reflection corresponds to the scattering from the SRO and LRO helix structure, respectively. Thus, as soon as the Bragg reflection is well described by the sum of Lorentzian and Gaussian functions with the same width and peak position one can separate the fractions of helical fluctuations and stable helical phase in the compound. The helical fluctuations have to exhibit a finite correlation length $2\pi / \kappa$ and lifetime $\tau$, which both are much smaller than the characteristic parameters for a LRO helical structure \cite{Grigoriev10PRB}.

\begin{figure}
\includegraphics[width=0.45\textwidth]{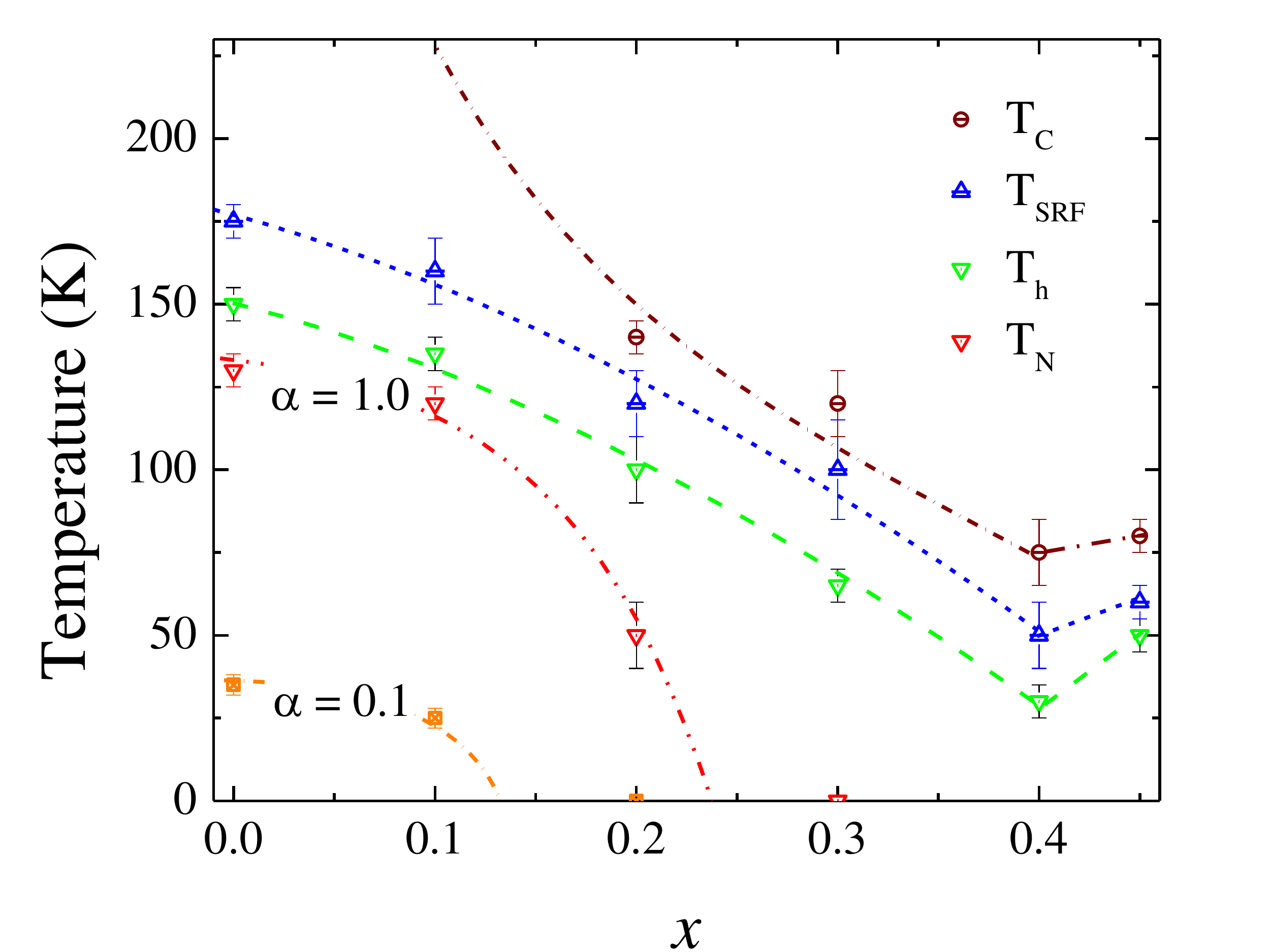}
\caption{(color online). $T-x$ phase diagram of the magnetic structure of Mn$_{1-x}$Co$_x$Ge compounds. $\alpha$ represents the fraction of the fluctuating spiral phase. The stable spiral phase (LRO) with $\alpha <0.1$ is the the ground state of pure MnGe in the left-down corner of the graph. The line $\alpha = 1.0$ defines the ($x-T$) point of the transition to the 100 \% fluctuating spiral state. This temperature is defined as $T_N$. The temperature $T_h$ determines the upper border of the fluctuating spiral phase. The temperature $T_{SRF}$ and $T_C$ defines the lower and upper border of the short-range ferromagnetic fluctuations respectively.}
\label{fig:Temps}
\end{figure}

 \begin{figure}
 \includegraphics[width=0.5\textwidth]{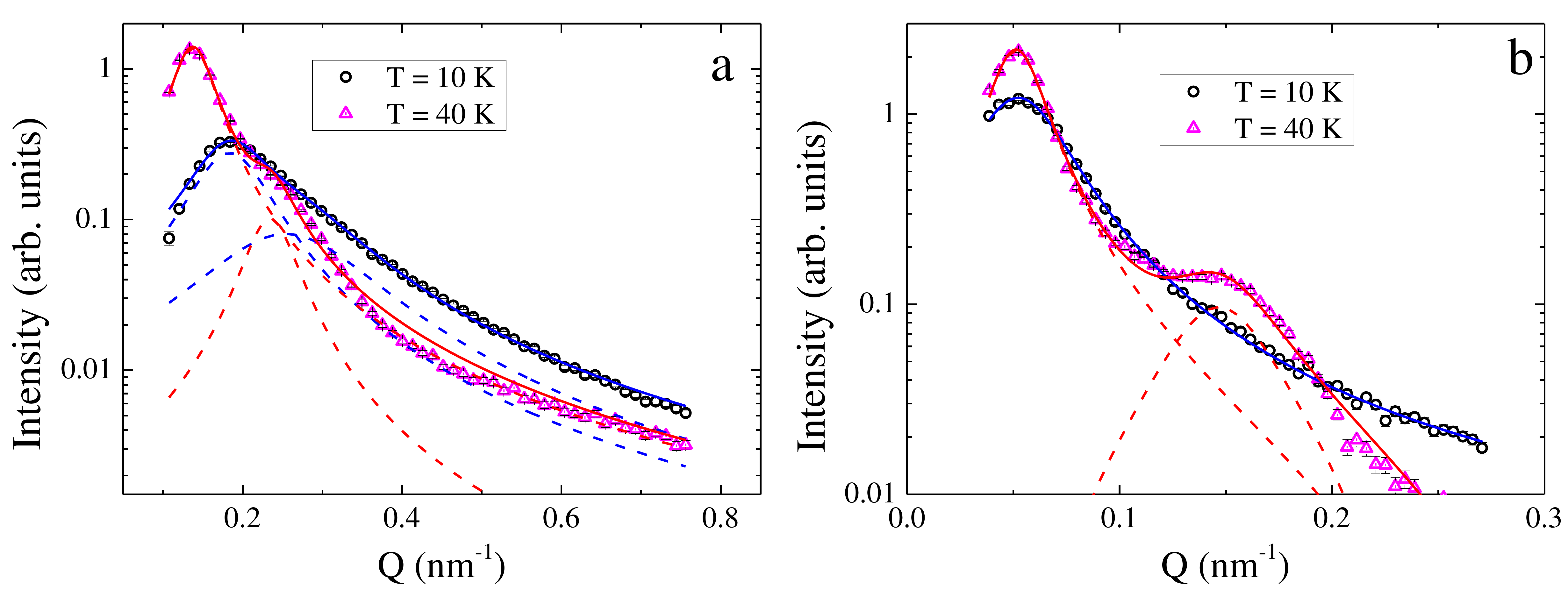}
 \caption{The momentum transfer dependence of the azimutally-averaged scattering intensity extracted from the SANS maps for $x = 0.5$ (a) and $x = 0.6$ (b) at $T = 10$ and 40 K. Solid lines correspond to thee best fit of data using two Lorentzian functions which are presented separately with dashed lines. The scattering profile for $x = 0.6$ at $T = 10$ K is well fitted with the single Lorentzian.}
 \label{fig:IvsQ}
 \end{figure}

 \begin{figure}
 \includegraphics[width=0.5\textwidth]{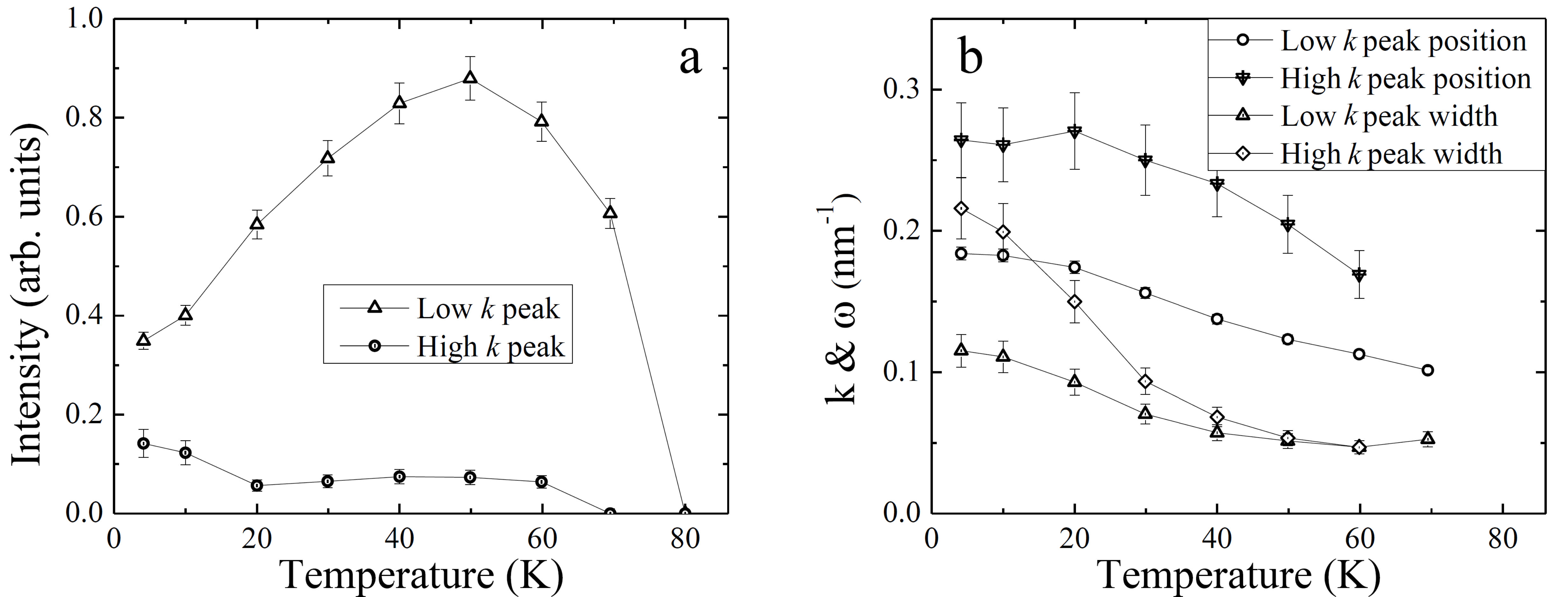}
 \caption{Temperature dependence of the integrated (a) intensity and (b) position and width of two different Bragg reflections observed with small angle neutron scattering from Mn$_{0.5}$Co$_{0.5}$Ge compound, at zero field.}
 \label{fig:50ikw}
 \end{figure}

 \begin{figure}
 \includegraphics[width=0.5\textwidth]{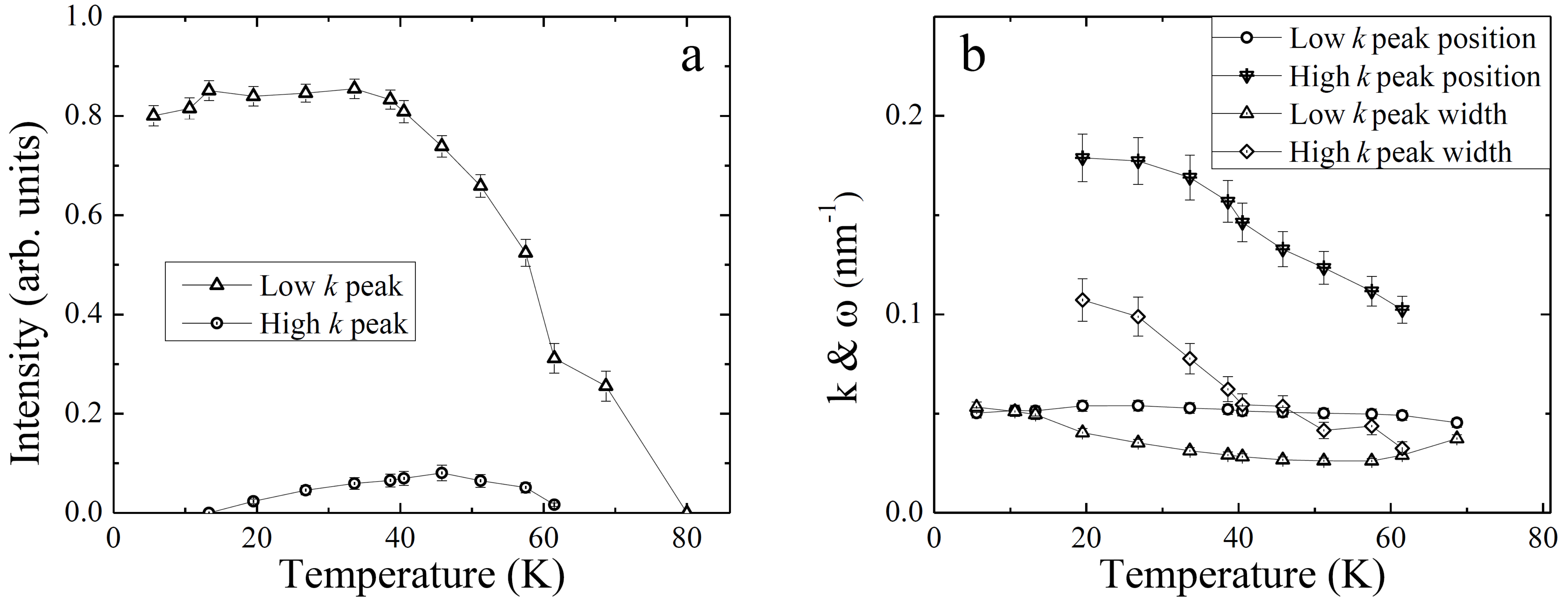}
 \caption{Temperature dependence of the integrated (a) intensity and (b) position and width of two different Bragg reflections observed with small angle neutron scattering from Mn$_{0.4}$Co$_{0.6}$Ge compound, at zero field.}
 \label{fig:60ikw}
 \end{figure}
 
As was found in \cite{Alt_14}, pure MnGe undergoes a series of transitions between different states of the magnetic structure. A stable helical structure was observed at low temperatures. With temperature increase the intensity of the Gaussian decreases down to zero while the intensity of the scattering coming from helical fluctuations increases and reaches its maximum value at $T_N = 130$ K. Peak corresponding to a helical structure can be observed below $T_{h} = 150$~K. A complex mixture of fluctuating spins, which can not be identified as a certain type of structure, is observed above $T_{h}$ and up to $T_{SRF} = 180$~K, where this mixture transforms into ferromagnetic fluctuations (defined as short-range ferromagnetic (SRF) state) with a characteristic size of not more than 2~nm. 

The analysis of the magnetic scattering profile of Mn$_{1-x}$Co$_x$Ge with $x < 0.45$ shows that the phase transition of the mixed compounds could be described in the same terms as for MnGe. As a result of the scattering profile evaluation, the ($T-x$) phase diagram of the magnetic structure of Mn$_{1-x}$Co$_x$Ge with $x < 0.5$ is plotted in Fig. \ref{fig:Temps}. It suggests that the 100\% fluctuating spiral state occurs in a large range of the $(T-x)$ phase diagram from $T_h$ down to the lower border marked as a line with $\alpha = 1.0$ (defining the temperature $T_N$). The temperature $T_h$ decreases smoothly with increasing $x$, while the temperature $T_N$ tends to zero with $x \rightarrow x_{c1} \approx 0.25$. For compounds with $x > 0.25$ the 100\% fluctuating spiral state ($\alpha = 1$) spreads out over the whole temperature range from the lowest measured temperatures up to $T_h$ (Fig. \ref{fig:Temps}).

The coexistence of the LRO and SRO of the spin helix structure at low temperatures is reflected in the non-zero value of $\alpha$. For pure MnGe $\alpha$ smoothly increases with temperature and is equal to 0.1 at $T \approx 35$ K. As it could be seen in Fig. \ref{fig:Temps} the fraction of SRO increases with $x$ and dominates completely at $x_{c1} \approx 0.25$. Together with the fact that the value of the helical wavevector decreases by one order of magnitude with $x$ within the range $x \in [0.25 \div 0.45]$ (Fig. \ref{fig:ivq}) the instability found at low temperatures should be related to the interplay between effective RKKY and DM interactions. Firstly, with increase of Co concentration in the compound up to $x_{c1} \approx 0.25$ the stable helical structure is completely disappeared meaning the reduction of the coupling between the second nearest neighbors. And secondly, further $x$ increase up to $x_{c2} \approx 0.45$ the $k$-value drops from 2.0 nm$^{-1}$ down to 0.2 nm$^{-1}$, that points out an increase of the DMI constant and change the main interaction that built the helical order from effective RKKY to DMI.


The Co substitution also leads to a decrease of the critical temperature $T_{SRF}$ and to the appearance of a phase where only paramagnetic scattering can be observed at temperatures above $T_C$ (Fig. \ref{fig:Temps}). This paramagnetic scattering is well-fitted with a single Lorentzian function centered at $Q = 0$. This paramagnetic state of the magnetic structure has so far not been found neither for pure MnGe nor for Fe-doped compounds and should appear at temperatures above 300 K \cite{Alt_14,Alt_16Pov,Alt_16PRB}.

\section{DMI-based helical structure}

The temperature evolution of the magnetic structure of Mn$_{1-x}$Co$_x$Ge with $0.45 < x < 0.9$ shows even more intriguing properties. As shown in Fig. \ref{fig:mapsx}e, the Bragg reflection for Mn$_{0.5}$Co$_{0.5}$Ge is strongly asymmetric. The border of the ring on the inner side ($Q < k$) is sharper than at higher $Q$. The momentum transfer dependence of the scattering intensity was extracted from the SANS maps, and its evolution with temperature has been investigated. The examples of the $I$ vs. $Q$ plots at temperatures $T = 10$ and 40~K and the data evaluation are presented in Fig. \ref{fig:IvsQ}a and b for Mn$_{1-x}$Co$_x$Ge with $x = 0.5$ and $0.6$, respectively. At temperatures $T < 70$ K and 20 K $< T < 60$ K for Mn$_{1-x}$Co$_x$Ge compounds with $x = 0.5$ and $0.6$, respectively, the profile consist of two different Lorentz peaks located at different $Q$-values. The presence of the second Lorentzian with a center position with higher value could be connected to the competition between DM interaction and effective RKKY-interaction that still plays an important role as it destabilizes the magnetic structure. The temperature evolution of the six fit parameters obtained from the data evaluation for Mn$_{1-x}$Co$_x$Ge with $x = 0.5$ and 0.6 is presented in Figs. \ref{fig:50ikw} and \ref{fig:60ikw}, respectively.

As it is shown in Fig. \ref{fig:50ikw} for Mn$_{0.5}$Co$_{0.5}$Ge, the increase of the temperature leads, firstly, to a loss of intensity of the peak with higher $k$ value together with an increase of intensity of the low-$k$ reflection (Fig. \ref{fig:50ikw}a), and, secondly, to a shift of both peak positions towards lower $Q$ values (Fig. \ref{fig:50ikw}b). It should be noted that the instrument resolution of SANS-1 was chosen as 20\% and is much smaller than the width of the reflections obtained from the data evaluation. The changes of the peak widths are strongly connected with the $k$-value (Fig. \ref{fig:50ikw}b). This means that the higher the temperature and the period of the fluctuating helix are, the more coherent is the magnetic structure. Thus, the increase of the correlation length of the magnetic structure with temperature is unexpectedly associated with a fragile balance between two coexisting interactions that generate the magnetic structure.

The analysis of the scattering profile of Mn$_{0.4}$Co$_{0.6}$Ge exhibits two remarkable features compared to the profile for $x=0.5$ (Fig. \ref{fig:60ikw}). First of all, the position of the main reflection, which has more intensity and lower $k$ value, is independent of the temperature. Its intensity decreases smoothly upon increasing temperature. This fact implies that the influence of the effective RKKY interaction, which is dominant in the compounds with lower Co concentration $x$, is already almost neglectable for $x = 0.6$. Nevertheless, the second feature of the magnetic structure is the appearance of a second reflection with temperature increase, which is connected to a non-vanishing influence of the still present RKKY interaction. Its position decreases rapidly, and the intensity has a maximum value at $T = 45$~K.

 \begin{figure}
 \includegraphics[width=0.45\textwidth]{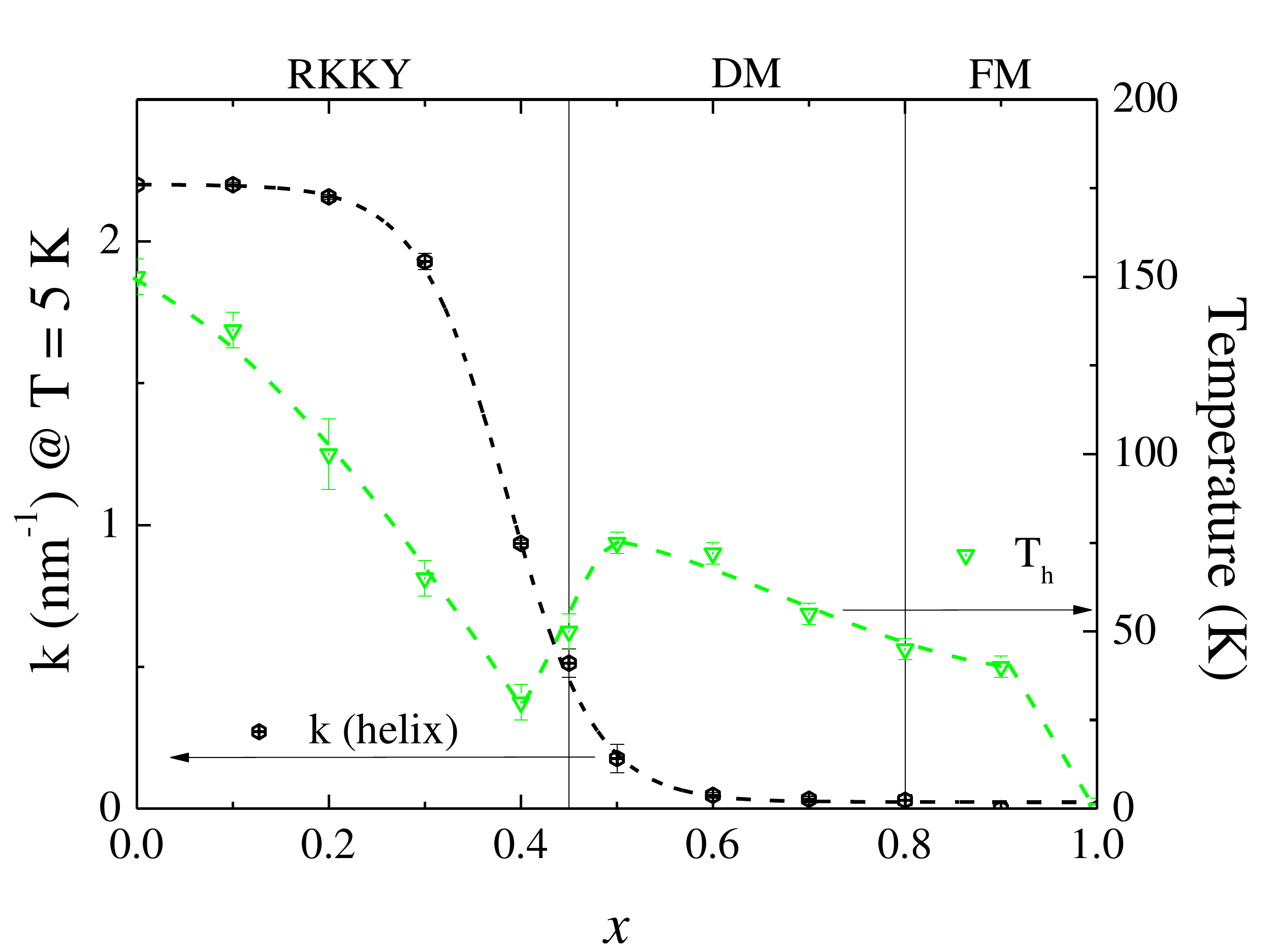}
 \caption{(color online) $x$-dependence of the helix wave vector value $k$ at temperature $T = 5$ K and of the temperature of the magnetic phase transition, $T_h$, extracted from the analysis of the temperature evolution of magnetic scattering profile. Lines are the guide for the eyes. Regions with $0.0 < x < 0.45$, $0.45 < x < 0.8$ and $0.8 < x < 1.0$ correspond to the RKKY-like helical, DMI-like helical and ferromagnetic-like structure of the Mn$_{1-x}$Co$_x$Ge system at $T = 5$ K.}
 \label{fig:TCkvx}
 \end{figure}

The experimental results for Mn$_{0.3}$Co$_{0.7}$Ge and Mn$_{0.2}$Co$_{0.8}$Ge do not show any scattering additional to the main reflections, the intensity of which decrease with rising temperature, whereas their position stays constant at $k = 0.03$ nm$^{-1}$ and 0.02 nm$^{-1}$ respectively.

The analysis of the scattering profile of the Mn$_{0.1}$Co$_{0.9}$Ge reveals a diffuse magnetic scattering centered at $k < 0.02$ nm$^{-1}$ meaning a ferromagnetic-like order of the compound at low temperatures. The intensity of this scattering decreases to zero with increasing temperature up to $T_{C} = 45$~K defining the ordering temperature for this compound. The true nature of the ferromagnetic order can be revealed taking the cubic anisotropy into account \cite{Grigoriev15PRB}. If the energy of the cubic anisotropy is high enough in compare to the energy of DMI, the ferromagnetic order is more preferable than the helical spin state.

\section{Conclusion}

Combining the experimental facts observed with SANS, the competition between the effective RKKY and the DM interaction, which is present in a large $x$ range up to $x < 0.8$ and generates the helical magnetic order, could be investigated. The DMI destabilizes the magnetic structure of Mn$_{1-x}$Co$_{x}$Ge with $x$ increase. The stable helical order disappears completely at $x_{c1} \approx 0.25$. Further increase of Co concentration leads to a dramatic decay of the wave vector $k$ at $x_{c2} \approx 0.45$ (Fig. \ref{fig:ivq}). This means that the DMI is the dominant interaction for compounds with $x > x_{c2}$. 

It is important to note that the critical concentration of Fe atoms in Mn$_{1-x}$Fe$_{x}$Ge compound needed to completely destabilize the helical structure is equal to $x_{c1} \approx 0.35$ while the decay of the wavevector value observed at the the same value of $x$ equal to $x_{c2} \approx 0.45$ \cite{Alt_16PRB}. Such a coincidence of critical values for Mn$_{1-x}$Fe$_{x}$Ge and Mn$_{1-x}$Co$_{x}$Ge compounds could not be connected to the changes of the electronic structure of MnGe system with Fe or Co replacement only because Co doping results in the increase of charge carrier density two times faster in compare to Fe replacement of Mn atoms. This fact could be assumed as another reason to consider the interactions between second nearest neighbors as a fundamental mechanism that creates the magnetic structure of pure MnGe.

SANS measurements of the magnetic structure of Mn$_{1-x}$Co$_{x}$Ge compounds also revealed that the effective RKKY interaction still remains notable and destabilizes the magnetic structure of Mn$_{1-x}$Co$_{x}$Ge with $0.45 < x < 0.6$. The DMI fully dominates over RKKY at low temperatures only for compounds with $x > 0.6$.

The resulting $x$-dependencies of wave vector $k$ and the critical temperature of the helices phase $T_h$ for compounds with $x < 0.8$ and $T_{C}$ for Mn$_{0.1}$Co$_{0.9}$Ge are presented in Fig. \ref{fig:TCkvx}. The relatively high value of the helical wavevector $k$ corresponds to a RKKY-like helical structure and is observed for $0.0 < x < 0.45$. The magnetic structure with $k \ll 1$ nm$^{-1}$ is ascribed to a DMI-like helical order at $x_{c2} < x < 0.8$. The critical temperatures obtained with Arrott plots for Mn$_{1-x}$Co$_{x}$Ge compounds with $x$ within the DMI region coincide with $T_h$, which is estimated from the treatment of the SANS data. The region in Fig. \ref{fig:TCkvx} with $0.8 < x < 1.0$ corresponds to compounds with non-zero critical temperature and undetectable helical wavevector, i.e. the ferromagnetic-like state of the magnetic structure of these compounds at $T = 5$ K.

In summary, a comprehensive small-angle neutron scattering study of the temperature evolution of Mn$_{1-x}$Co$_x$Ge allows one to consider the RKKY as the fundamental interaction for the helical structure in MnGe. It can be concluded that an order-disorder phase transition takes place with increasing $x$ at $x_{c1} \approx 0.25$ caused by the modification of the effective Ruderman-Kittel-Kasuya-Yosida exchange interaction within the Heisenberg model of magnetism. The DMI can be considered as an instrument for destabilization of the ordered helical structure with $x$ or $T$, despite the fact that all Mn$_{1-x}$Co$_x$Ge compounds crystallize in the B20 structure. With further increase of $x$ ($x > x_{c2} \approx 0.45$) these two interactions generating the magnetic structure compete with each other in the sense that the RKKY interaction destabilizes the DMI-based helical order until $x$ exceeds the value of 0.6. After that the effective RKKY can be neglected, and only the DMI and the cubic anisotropy determine the magnetic order in Mn$_{1-x}$Co$_x$Ge as being ferromagnetic or helical depending on the ratio between them.


The work was supported by the Russian Foundation of Basic Research (Grant No 14-22-01073, 14-02-00001) and the special program of the Department of Physical Science, Russian Academy of Sciences.

\end{document}